# Edge-induced Schottky barrier modulation at metal contacts to exfoliated molybdenum disulfide flakes

Ryo Nouchi[a)]

*Nanoscience and Nanotechnology Research Center, Osaka Prefecture University, Sakai 599-8570, Japan*

ABSTRACT. Ultrathin two-dimensional semiconductors obtained from layered transition-metal dichalcogenides such as molybdenum disulfide ($MoS_2$) are promising for ultimately scaled transistors beyond Si. Although the shortening of the semiconductor channel is widely studied, the narrowing of the channel, which should also be important for scaling down the transistor, has been examined to a lesser degree thus far. In this study, the impact of narrowing on mechanically exfoliated $MoS_2$ flakes was investigated according to the channel-width-dependent Schottky barrier heights at Cr/Au contacts. Narrower channels were found to possess a higher Schottky barrier height, which is ascribed to the edge-induced band bending in $MoS_2$. The higher barrier heights degrade the transistor performance as a higher electrode-contact resistance. Theoretical analyses based on Poisson's equation showed that the edge-induced effect can be alleviated by a high dopant impurity concentration, but this strategy should be limited to channel widths of roughly 0.7 μm because of the impurity-induced charge-carrier mobility degradation. Therefore, proper termination of the dangling bonds at the edges should be necessary for aggressive scaling with layered semiconductors.

[a] E-mail: r-nouchi@21c.osakafu-u.ac.jp

## I. INTRODUCTION

The continuous shrinking of Si-based transistors has led to cost reduction, performance improvement, and reduced power consumption of semiconductor devices. However, it is believed that the advancement of the miniaturization of the transistors is reaching its limit.[1] A main factor determining the limit is the short-channel effect, which increases the off-state leakage current.[2] The leakage increases the power consumption and deteriorates the switching performance. The deteriorative short-channel effect is known to be alleviated by the thinning of the semiconductor channel body.[3] Thus, ultrathin sheets obtained by the exfoliation of layered semiconductors such as group 6 transition-metal dichalcogenides (TMDCs) are now considered as a promising post-Si material.[4,5]

Another critical obstacle of device shrinking is the electrode-contact resistance, which exists parasitically at interfaces between metallic electrodes and a semiconductor channel.[1,6] The electrode-contact resistance dominates the electrical resistance of short-channel devices and hampers the efficient electron transport through the devices. To reduce the electrode–semiconductor interfacial resistance, the energy barrier for charge-carrier injection from the electrode into the semiconductor (a Schottky barrier) should be controlled. The controllability of the Schottky barrier height closely relates to the density of electrode–semiconductor interface states.[7,8] For three-dimensional (3D) semiconductors such as Si, the density of the interface states is considerably high because of the electronic states arising from surface dangling bonds, which yields an almost complete loss of the controllability of the Schottky barrier height.[9,10] In contrast, the loss of the controllability, which is known as the Fermi-level pinning, is less distinct for two-dimensional (2D) TMDC sheets such as $WSe_2$ (Refs. 11, 12) because a number of dangling bonds are expected to be small owing to the van der Waals bonded layered crystal structure of TMDCs.

However, even in layered crystals, the edges break the translational symmetry of the crystal structure, and thus dangling bonds are unavoidable at the edges. A conventional scaling law for the device shrinking (Dennard scaling[13]) indicates that the miniaturization of transistors requires a reduction in the semiconductor channel width. Considering a very narrow channel of a tiny transistor, the contribution of the channel edges should be significantly large. For 3D semiconductors, dangling bonds are present also in the channel plane, and thus the channel edge is never a special point. In contrast, for the 2D sheets, the edge is the only point where high-density dangling bonds are expected. Despite its importance, the effect of the TMDC edge on the electrode contacts remains largely elusive.

In this study, a significant effect of the edges of semiconducting TMDC sheets on the electrode contact was unveiled by investigating the channel-width dependence of the Schottky barrier height at Cr/Au electrode-exfoliated $MoS_2$ flake interfaces. The Schottky barrier height was found to be higher as the channel width at the injection contact (source contact) decreased, which can be understood by the Fermi-level pinning to (the charge neutrality level[14] of) the edge states. This is the direct observation of the deteriorative effect of the edges on the electrode contact, and it indicates the need for the proper termination of dangling bonds at the edges of semiconducting 2D sheets.

## II. EXPERIMENTAL

A highly doped Si wafer with a 300-nm-thick thermal oxide layer on top of it was used as a substrate. The substrate was cleaned with acetone and isopropanol by using an ultrasonic bath, followed by oxygen plasma cleaning. Immediately after the plasma cleaning, MoS$_2$ flakes were formed on the substrate by mechanical exfoliation from a natural MoS$_2$ crystal (SPI Supplies) in ambient air. Multiple Au electrodes with a 1-nm-thick Cr adhesion layer were fabricated on the flakes with a triangular shape by conventional electron-beam lithography processes. The inter-electrode spacing (channel length) was set to 1 μm. The temperature dependence of the FET characteristics was measured by a low-temperature vacuum prober (<10$^{-3}$ Pa) equipped with a semiconductor device parameter analyzer (Keysight, BA1500A) under a dark condition. The thicknesses of the flakes were determined by an atomic force microscope (SII, SPA-400).

## III. RESULTS

Figure 1(a) shows an optical micrograph of sample Cr/Au-contacted MoS$_2$ field-effect transistors (FETs) on a highly-doped Si wafer with a 300-nm-thick thermal oxide layer. The MoS$_2$ flakes were used as-exfoliated in order to avoid introducing damage by micropatterning with energetic particles. By selecting flakes with a naturally formed triangular shape, multiple transistors with varied channel widths can be fabricated on a single flake. The channel length was set to 1 μm. As schematically depicted in Fig. 1(b), lithographically patterned Au electrodes with a Cr adhesion layer were fabricated to be placed across the both edges. MoS$_2$ transistors are known as Schottky-barrier transistors, and the charge injection from the source electrode limits the transistor performance.[15] Therefore, the channel width at the source contact is considered below.

The drain current $I_{DS}$ of such charge-injection-limited devices is expressed as[16]

$$I_{DS} = AA_{2D}^{*}T^{3/2}\exp\left(-\frac{q}{k_{B}T}\left(\Phi_{B}-\frac{V_{DS}}{n}\right)\right), \tag{1}$$

$$\ln\frac{I_{DS}}{T^{3/2}} = -\frac{q}{k_{B}T}\left(\Phi_{B}-\frac{V_{DS}}{n}\right)+\ln\left(AA_{2D}^{*}\right), \tag{2}$$

where $A$ is the effective area of the source contact, $A_{2D}^{*}$ is the 2D equivalent Richardson constant, $T$ is the absolute temperature, $q$ is the electronic charge, $k_B$ is the Boltzmann constant, $\Phi_B$ is the zero-bias Schottky barrier height, $V_{DS}$ is the drain voltage, and $n$ is the ideality factor. From this expression, $\Phi_B$ can be obtained by first extracting the slope of $\ln(I_{DS}/T^{3/2})$–$T^{-1}$ curves measured with various $V_{DS}$ values and then extrapolating to zero $V_{DS}$.

Figure 2 shows a procedure for extracting the Schottky barrier heights. The data are based on the FET characteristics of a thin MoS$_2$ flake with a thickness of 23.5 nm. The channel width at the source end was 2.3 μm for this example device. Fig. 2(a) shows the temperature dependence of the transfer characteristics measured with $V_{DS} = 0.2$ V. The horizontal axis is the relative $V_G$ value from the threshold voltage $V_T$. By the linear fitting of $I_{DS}$–$V_G$ curves around a maximum-transconductance point, the $V_T$ values were extracted from the x-intercept of the fitted line (here, $V_G$ is on the x axis), as is routinely performed in analyses of the linear region of FETs.[17] From the slope of the $\ln(I_{DS}/T^{3/2})$–$T^{-1}$ curve at each $V_G$, the $V_G$ dependence of the barrier height ($\Phi_B - V_{DS}/n$) was obtained as shown in Fig. 2(b). The error bars show the standard error of the least-squares fit. Figure 2(c) shows the $V_{DS}$ dependence of the barrier height at $V_G = V_T - 10$ (V). By extrapolating to $V_{DS} = 0$ at each $V_G$, the zero-bias barrier height $\Phi_B$ was obtained as shown in Fig. 3(a).

Figure 3(a) is the $V_G$ dependence of the zero-bias Schottky barrier height extracted from the data based on the FET characteristics of a thin MoS$_2$ flake with a thickness of 23.5 nm, which can be explained as depicted in Fig. 3(b). In region I, the application of $V_G$ induced the upward bending of the electronic bands in MoS$_2$, leading to a continuous increase in $\Phi_B$. The charge-carrier injection through the metal contact in region I is ascribed to thermionic emission over the Schottky barrier. In regions II and III, the application of $V_G$ induced the downward bending of the bands. In region II, the charge injection is ascribed to the thermionic emission over the barrier. To the contrary, the further increase in $V_G$ in region III induced the narrowing of the Schottky barrier, causing tunneling carrier injection. Thus, the extracted $\Phi_B$ values in region III ($V_G > V_T$) were considerably low. The behavior, especially in region III, is different from previous reports with single-layer[18] and multilayer[19] MoS$_2$ flakes, which should be attributed to the lower barrier heights in the present study. The lower barrier heights might be due to combined effects of a larger thickness[20] and rather low work function of the adhesion layer metal (4.5 eV for polycrystalline Cr[21]). The lower barrier can lead to carrier transport limited not only by the carrier injection through the metal contact, but also by the band-like transport within the MoS$_2$ flakes. The band-like transport appears upon the channel formation, which might have caused the sudden decrease in the barrier height in region III ($V_G > V_T$). According to the aforementioned considerations, the Schottky barrier height under flat-band condition was acquired, as indicated by the gray line in Fig. 3(a).

Figure 4(a) compiles the flat-band zero-bias Schottky barrier heights extracted for Cr/Au contacts to several exfoliated MoS$_2$ flakes. The vertical axis was set to the flat-band barrier height relative to a wide-channel value. For sample #1, the four values from the right endpoint do not show a clear tendency, and the average value of these four values is taken as the zero point.

For sample #2, all the data points are almost featureless, and the average of all the values is taken as the zero point. For samples #3 and #4, the data points show a monotonic increase as the channel width decreased, and the right endpoints are taken as the zero point. The wide-channel value should be the same as the flat-band $\Phi_B$ values for in-plane metal contacts without covering the edges. For metal contacts to $MoS_2$, strong Fermi-level pinning is known to occur[19,22] owing to in-plane sulfur vacancies.[23,24] However, the in-plane defect density of the as-exfoliated flakes of a natural $MoS_2$ crystal has been elucidated, showing a flake-to-flake variation up to 8%.[25] This variation leads to different degrees of the Fermi-level pinning, yielding a flake-to-flake variation of $\Phi_B$. To compile the data from different samples without the influence of the variation of in-plane defect densities, flat-band zero-bias barrier heights relative to a wide-channel value are used in Fig. 4(a).

The dataset in Fig. 4(a) shows a rather large noise floor of roughly 40 meV, which might be attributable to various facts. The extraction method of the Schottky barrier height relies on the measurement of $I_{DS}$–$T$ characteristics, and thus the energy resolution might be limited by the thermal energy ($k_B T$). In addition, the very thin Cr layer (1 nm thick) forms a discontinuous/cluster film on $MoS_2$;[26] the electrode–$MoS_2$ interface should consist of two contact metals (Cr and Au). Thus, the morphology should vary from point to point within the electrode. However, the effect of the position dependent morphology on the extracted Schottky barrier height is expected to be small because the total area covered by the Cr may not change so much. The Schottky barrier inhomogeneity caused by the partial coverage of the very thin Cr film should be treated as the parallel connection of the current path with different barrier heights. Thus, the extracted Schottky barrier height is not influenced by the difference in the local morphology film, but by the total area covered by Cr, leading to a limited influence of the very thin nature of the Cr adhesion layer. Although these noise sources might have contributed to the rather noisy dataset in Fig. 4(a), a feature in the narrow-channel region of the plot is considered to surpass the noise floor.

The data compiled in Fig. 4(a) indicate that the flat-band Schottky barrier height increased when the channel width at the source end decreased below ~4 μm. The increase in the flat-band barrier height can be understood according to the band bending in MoS$_2$ due to the presence of edges. In the charge-neutral condition (the left panel of Fig. 4(b)), the Fermi level at the edge $E_{Fe}$ differed from that of the point far from the edge $E_{Fb}$. The $E_{Fe}$ is generally located around the mid gap in analogy with the surface states of 3D semiconductors,[27,28] and was indeed found to be located around the mid gap for single layer MoS$_2$.[29] On the other hand, the $E_{Fb}$ of MoS$_2$ is located close to the conduction-band minimum owing to the Fermi-level pinning to the electronic states originating from in-plane sulfur vacancies,[30,31] as previously observed.[19,22] This is corroborated in the present study by the very low (< 100 meV) value of the $\Phi_B$ values shown in Fig. 3(a). In the thermal-equilibrium condition (the right panel of Fig. 4(b)), the Fermi-level alignment caused upward band bending toward the edge. Although the Fermi-level pinning due to in-plane defects in MoS$_2$ is known to be strong, the presence of the edge-induced band bending indicates that the pinning strength due to the edges was stronger than that due to the in-plane defects. Thus, the edge-induced band bending of MoS$_2$ increases $\Phi_B$ as the channel becomes narrower.

If $I_{DS}$ is fully characterized by Eq. (1), $I_{DS}/W$ at low $V_{DS}$ should be determined solely by $\Phi_B$. Thus, for example, the decrease in $\Phi_B$ of 50 meV upon the increase in cannel width from 1.3 to 4.0 μm (see sample #3 in Fig. 4(a)) should lead to a 5.7-fold increase in $I_{DS}/W$. However, transfer characteristics of these two devices measured with $V_{DS}$ of 0.1 V and at 333.3 K display only ~2-fold increase as shown in Fig. 5. This can be attributed mainly to the so-called fringing field.[32] Because of the fringing field, the effective strength of the gate electric field for narrow-channel devices becomes stronger than the value for an ideal parallel-plate capacitor. As a result, the increase in $I_{DS}/W$ caused by the decrease in $\Phi_B$ is compensated by the effect of the fringing field. In addition, the transport channel through metallic edge states[33] might also compensate it. Although quantitative formulations of these effects on MoS$_2$ FETs are unknown at present, these two effects clearly weaken the increase in $I_{DS}/W$.

## IV. DISCUSSION

The edge-induced band bending can be formulated by solving the one-dimensional Poisson's equation in the depletion region adjacent to the edge:

$$\frac{d^2\phi(y)}{dy^2} = -\frac{qN_d}{\varepsilon_r\varepsilon_0}, \tag{3}$$

where $\phi$ is the electrostatic potential, $y$ is the distance from the edge toward the channel center, $q$ is the electric charge, $N_d$ is the concentration of ionized donors, $\varepsilon_r$ is the dielectric constant, and $\varepsilon_0$ is the permittivity of the vacuum. By integrating once under the boundary condition of zero electric field at the channel center, i.e., $d\phi(W_D)/dy = 0$, where $W_D$ is the depletion width due to the edge-induced band bending, we obtain

$$\frac{d\phi(y)}{dy} = -\frac{qN_d}{\varepsilon_r\varepsilon_0}(y-W_D). \tag{4}$$

By setting the channel center to the zero point of the potential, i.e., $\phi(W_D) = 0$, the electrostatic potential can be obtained as

$$\phi(y) = -\frac{qN_d}{2\varepsilon_r\varepsilon_0}\left(y^2 - 2W_D y + W_D^2\right). \tag{5}$$

The flat-band zero-bias Schottky barrier heights extracted in Fig. 4(a) should have been affected by the edge-induced band bending expressed by Eq. (5). The band bending occurred along the channel-width direction, which was perpendicular to the flow of the charge carriers. Therefore, the measured $I_{DS}$ should be expressed by the parallel connection of channels with different $\Phi_B$ values. The expression for $I_{DS}$ differs from Eq. (1) in the main text and instead becomes

$$I_{DS} = AA_{2D}^* T^{3/2} \exp\left(\frac{qV_{DS}}{nk_B T}\right)\frac{1}{W}\int_0^W \exp\left(-\frac{q\Phi_B(y)}{k_B T}\right)dy, \tag{6}$$

where $W$ is the channel width at the source end. By comparing Eqs. (1,6), the $\Phi_B$ values extracted from experimentally obtained data, $\Phi_B^{exp}$, should be expressed as

$$\Phi_B^{exp} = \frac{k_B T}{q}\ln\left[\frac{1}{W}\int_0^W \exp\left(-\frac{q\Phi_B(y)}{k_B T}\right)dy\right]. \tag{7}$$

Here, the $y$-dependent $\Phi_B$ can be expressed as

$$\Phi_B(y) = \Phi_{B\infty} - \phi(y) \ (0 \leq y \leq W_D, W - W_D \leq y \leq W), \tag{8}$$

$$\Phi_B(y) = \Phi_{B\infty} \ (W_D \leq y \leq W - W_D), \tag{9}$$

where $\Phi_{B\infty}$ is the $\Phi_B$ value for the wide-channel limit, and the negative sign in front of $\phi$ reflects the electron conduction.

The aforementioned expressions, which were derived by including the edge-induced band bending, were used to explain the features of the extracted flat-band Schottky barrier heights, as shown in Fig. 6. The $\varepsilon_r$ of the rather thick MoS$_2$ flakes used in this study was ~10.5,[34] and $N_d$ and $W_D$ are unknown parameters. The parameter $W_D$ can be related to $N_d$ by introducing the built-in potential, $\phi_{bi}$, as

$$\phi_{bi} = |\phi(0)| = \frac{qN_d W_D^2}{2\varepsilon_r \varepsilon_0}, \tag{10}$$

$$W_D = \sqrt{\frac{2\varepsilon_r \varepsilon_0 \phi_{bi}}{qN_d}}. \tag{11}$$

Here, $\phi_{bi}$ under zero bias is expressed by the energy difference in the Fermi levels: $|E_{Fb} - E_{Fe}|$. At the edge, $E_{Fe}$ is known to be located around the mid gap for single layer MoS$_2$,[29] which is consistent with the calculation result that the charge-neutrality level of the surface states of 3D semiconductors is located slightly below the mid gap.[27] On the other hand, the in-plane Fermi level, $E_{Fb}$, should be close to the conduction-band minimum, as previously reported.[19,22,30,31] This is corroborated in the present study by the very low (< 100 meV) value of the $\Phi_B$ values shown in Fig. 4(a). According to these considerations, $q\phi_{bi}$ can be approximated by half of the bandgap, and thus $\phi_{bi} \approx 0.65$ V.[35] Therefore, only $N_d$ remains as an unknown parameter.

Figure 6(a) shows $\phi(y)$ calculated with various $N_d$ values. A higher dopant concentration naturally caused a steeper change in $\phi(y)$ because of the more efficient screening by free electrons, which yielded a smaller $W_D$, as shown in Fig. 6(b). Considering the band bending from the both edges, the whole channel was affected by the edges when $W < 2W_D$, where $W$ is the channel width at the source end. The measured $I_{DS}$ should be expressed by the parallel connection of channels with different $\Phi_B$ values caused by the edge-induced band bending, and the $\Phi_B$ values extracted from experimentally obtained data, $\Phi_B^{exp}$, becomes $y$-dependent as expressed by Eqs. (6–9). A variation in $\Phi_B^{exp}$ due to the edge-induced band bending, $\Delta\Phi_B^{exp} \equiv \Phi_B^{exp} - \Phi_{B\infty}$, is shown in Fig. 6(c). The variable $\Phi_B^{exp}$ was determined by the integration process expressed by Eq. (7), and was thus considerably affected even when $W > 2W_D$. The calculated $\Phi_B^{exp}$ naturally changed more steeply with a higher $N_d$.

A common feature in Figs. 6(a–c) is that a higher $N_d$ causes a steeper change within a shorter distance from the edges. Thus, the effect of the edge-induced Fermi-level pinning was alleviated by higher dopant concentrations. Figure 6(d) shows the threshold channel width, $W_T$, which was calculated to be equal to $W$ for $\Delta\Phi_B^{exp} = k_BT$ at 300 K (~26 meV). The error bars arise from the minimum step size used for the numerical calculation. When $W < W_T$, the device performance was considerably degraded because the increase in $\Phi_B^{exp}$ was higher than the thermal energy. Even with a very high $N_d$ of $10^{18}$ cm$^{-3}$, $W_T$ was above 50 nm; $N_d > 3 \times 10^{19}$ cm$^{-3}$ was required to obtain $W_T < 10$ nm.

However, a higher $N_d$ enhances the ionized-impurity scattering of the charge carriers, which degrades the transistor performance. In the case of 3D semiconductors, electron and hole mobilities at 300 K start to decrease when the dopant concentration becomes higher than ~$10^{16}$ cm$^{-3}$.[36] Presently, this critical dopant concentration is unknown for 2D semiconductors such as MoS$_2$. By employing the same $N_d$ as the critical value, the upper limit of $W_T$ without considerable degradation of the charge-carrier mobilities is determined to be 710 nm from Fig.

6(d). Therefore, it is impractical to achieve ultimate miniaturization solely with the high-$N_d$ strategy. and the proper termination of dangling bonds at the edges is necessary to lift the edge-induced Fermi-level pinning.

The effect of the edges of TMDC sheets on metal contacts has been comprehensively studied by *ab initio* density-functional theory calculations.[37] In the case of an edge contact, in-gap metallic electronic states have been found to be formed by covalent bonds between metal atoms and TMDC edge atoms, which results in zero charge-injection barrier height. However, the zero barrier height holds for carrier injection to the in-gap states localized at the edges, and subsequent carrier propagation to the channel unperturbed by the metal contact should be largely prevented by finite injection barriers. The zero barrier height can also hold for charge injection to metallic edge states[33] of bare TMDC edges. However, a microscopic fluctuation of the edge structure naturally exists in the as-exfoliated flakes, which easily prevents the perfect connection of the metallic edge states from the source to drain (the channel length is 1 μm). Thus, the edge states as a transport channel can be disregarded here. Although the electrode metal might contact the edges of only top several layers due to the rather thick $MoS_2$ flakes used in the present study,[38] this structural complexity of the electrodes may not affect the extracted heights of the overall Schottky barriers.

The relative $\Phi_B$ values extracted from the experimental data (Fig. 4(a)) were directly compared with the calculated $\Phi_B^{exp}$ shown in Fig. 6(c). We estimated $N_d$ from the comparison. The $MoS_2$ flakes used in this study were mechanically exfoliated from a natural $MoS_2$ crystal and thus contained various impurity atoms.[39, 40] In addition, they had sulfur vacancies.[24,40] The densities of the impurity atoms and vacancies exhibit a flake-to-flake variation,[25] which caused a variation in $N_d$. By comparing Figs. 4(a) and 6(c), the variation in the $N_d$ of the flakes used in this study can be determined to range from $5 \times 10^{14}$ to $3 \times 10^{15}$ cm$^{-3}$. These values are close to the reported value of $3.61 \times 10^{10}$ cm$^{-2}$ for an 80-nm-thick MoS$_2$ flake,[41] corresponding to $4.5 \times 10^{15}$ cm$^{-3}$.

Similar $W$-dependent features in MoS$_2$ FETs have been reported by Liu *et al*.[42] They have found that the threshold voltage increased as $W$ became narrower than ~0.2 μm, being attributed to the edge depletion. This result is explainable by the edge-induced band bending described in the present study; the increase in $\Phi_B^{exp}$ in the narrower device should cause the depletion of the channel. However, the $W$ range for the threshold voltage shift is different from that for the increase in the flat-band Schottky barrier height (Fig. 4(a)). This difference might be attributed to the difference in $N_d$; the $N_d$ value for the flakes in Ref. 42 can be estimated to be ~$10^{17}$ cm$^{-3}$. This discrepancy might be again attributable to the large flake-to-flake variation.[25]

## V. CONCLUSION

The effects of the edges of layered 2D semiconductors on metal contacts were studied by examining the channel-width dependent Schottky barrier heights at Cr/Au contacts to exfoliated MoS$_2$ flakes. The Schottky barrier height increased as the MoS$_2$ channel width decreased. This behavior is explained by the edge-induced band bending in MoS$_2$, which indicates that the strength of the Fermi-level pinning due to the edges was stronger than that due to the in-plane defects in MoS$_2$. From simple theoretical analyses based on Poisson's equation, high ionized dopant impurity concentrations can alleviate the edge-induced effect. However, the narrowing with this high-dopant strategy was found to be limited to roughly ~0.7 μm, and an attempt to achieve further narrowing with higher dopant concentrations led to a significant degradation in the charge-carrier mobility because of the enhanced ionized impurity scattering. A comparison between the analyses of the experimentally extracted data indicated an ionized-impurity concentration within the range of $5 \times 10^{14}$ to $3 \times 10^{15}$ cm$^{-3}$ in our samples exfoliated from a natural MoS$_2$ crystal, which is close to the previously reported value.

The narrowing of semiconductor channels, as well as shortening, is critically important for the miniaturization of electronic devices. The present study unveiled the significance of the effect of edges on Schottky barrier heights of narrow-channel devices. Although the thicknesses of the MoS$_2$ flakes used in this study are rather thick (> 14 nm), the conclusions drawn from the present results should also be applicable to thinner flakes with edges. The proper termination of the dangling bonds at the edges of layered semiconductors should be necessary for aggressive narrowing without the degradation of the transistor performance.

**ACKNOWLEDGMENTS**

This work was supported in part by the Special Coordination Funds for Promoting Science and Technology and a Grant-in-Aid for Scientific Research on Innovative Areas (No. 26107531) from the Ministry of Education, Culture, Sports, Science and Technology of Japan, as well as by the Tokuyama Science Foundation.

Figures:

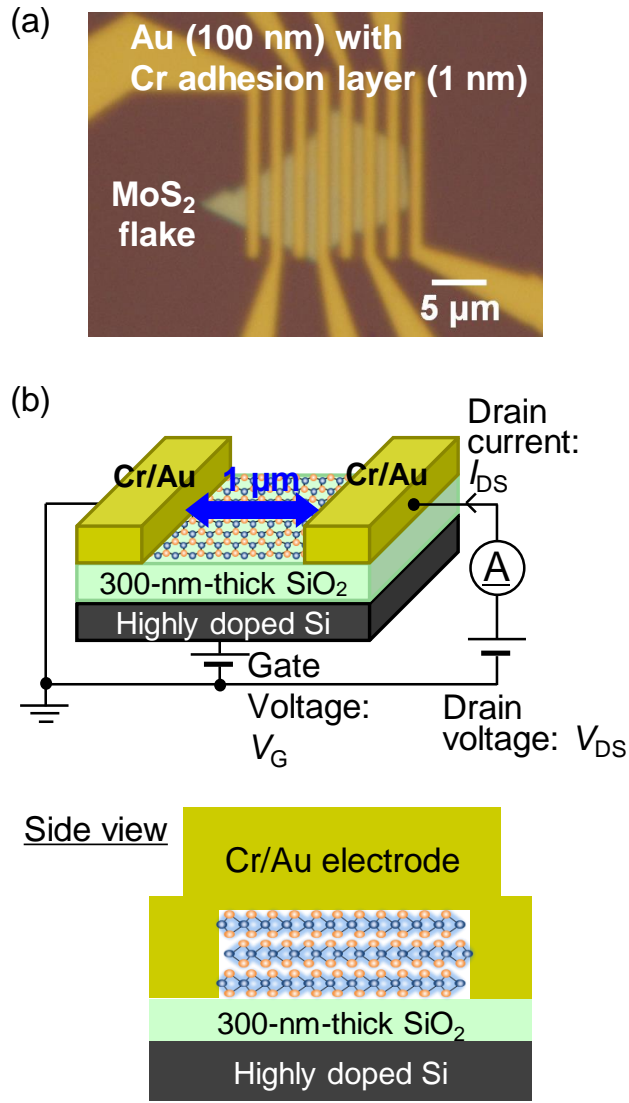

FIG. 1. Structure of MoS$_2$ FETs fabricated in this study. (a) Optical micrograph of multiple Cr/Au electrodes on MoS$_2$ flakes with different channel widths. MoS$_2$ flakes with a naturally formed triangular shape were used as-exfoliated in order to avoid the possible damage introduced by micro-patterning with energetic particles. The channel length was set to 1 μm. (b) Schematic (top) and side view (bottom) of the MoS$_2$ transistors. The electrodes were fabricated to be placed across the both edges.

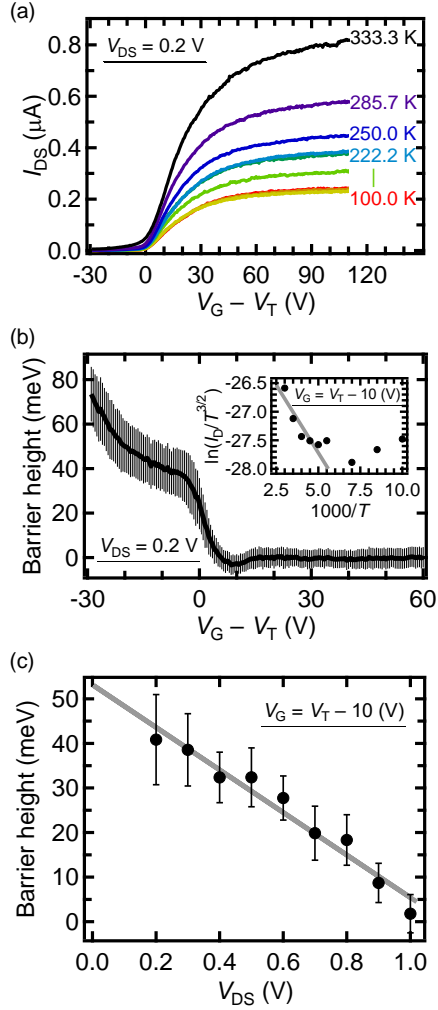

FIG. 2. Procedure for extracting the Schottky barrier heights. The data are based on the FET characteristics of a thin $MoS_2$ flake with a thickness of 23.5 nm and a channel width of 2.3 μm at the source end. (a) Temperature dependence of the transfer characteristics measured with the drain voltage $V_{DS} = 0.2$ V. The horizontal axis shows the relative gate voltage $V_G$ from the threshold voltage $V_T$. (b) $V_G$ dependence of the barrier height $(\Phi_B - V_{DS}/n)$ obtained from the slope of the $\ln(I_{DS}/T^{3/2})$–$T^{-1}$ curve at each $V_G$. The error bars show the standard error of the least-squares fit. The inset shows the $\ln(I_{DS}/T^{3/2})$–$T^{-1}$ curve at $V_G = V_T - 10$ (V). The fitting range was $3 \leq 1000/T \leq 5$, where the contribution from the tunneling injection was insignificant. (c) $V_{DS}$ dependence of the barrier height at $V_G = V_T - 10$ (V).

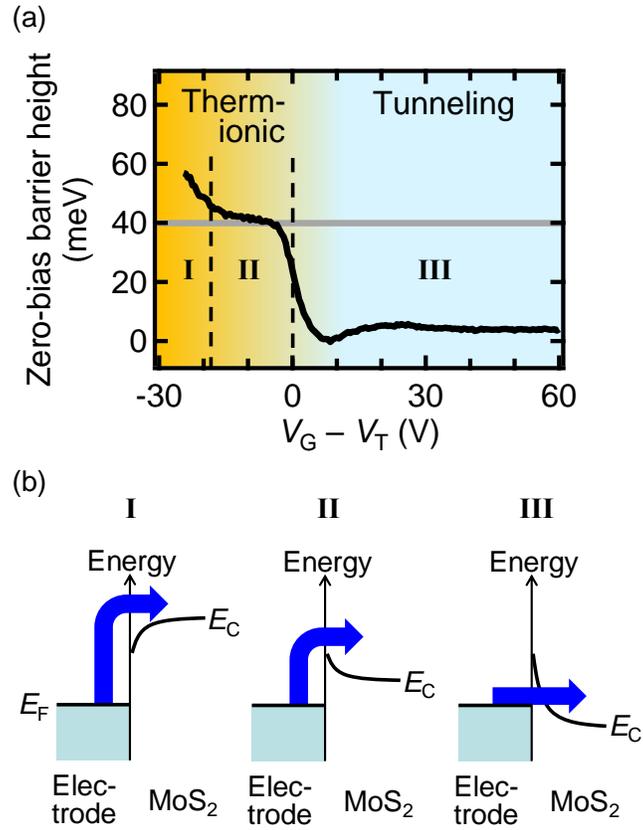

FIG. 3. Extracted zero-bias Schottky barrier heights at the Cr/Au–MoS$_2$ interface. The MoS$_2$ flake used to extract this data is identical to what was used in Fig. 2. (a) $V_G$ dependence of the zero-bias Schottky barrier height obtained by extrapolating to $V_{DS} = 0$ at each $V_G$. The gray line indicates the Schottky barrier height under the flat-band condition. (b) Schematic band diagrams for stages I–III in (a). $E_F$ and $E_C$ denote the Fermi level and the conduction-band minimum, respectively.

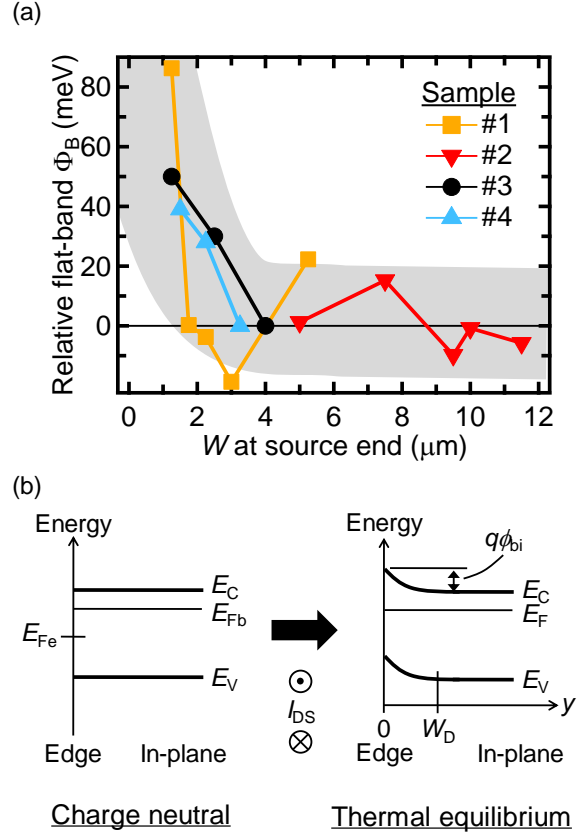

FIG. 4. Channel-width dependence of extracted flat-band Schottky barrier heights at the Cr/Au–MoS$_2$ interface. (a) Flat-band zero-bias Schottky barrier heights extracted for Cr/Au contacts to four exfoliated MoS$_2$ flakes. To compile the data from different samples without the influence of the variation of in-plane defect densities, the flat-band zero-bias barrier heights relative to a wide-channel value were used. The wide-channel values are explained in detail in the main text. The thicknesses of the four samples were 14.1, 28.5, 23.5, and 29.1 nm for samples #1–4, respectively. (b) Schematic band diagram of MoS$_2$ in the charge-neutral condition (left) and the thermal-equilibrium condition (right). $E_{Fe}$ and $E_{Fb}$ denote the Fermi level at the edge and that at a point far from the edge, respectively. $E_C$, $E_V$, $\phi_{bi}$, and $W_D$ are the conduction-band minimum, valence-band maximum, built-in potential, and depletion width, respectively. It should be noted that the current flow direction is perpendicular to the paper.

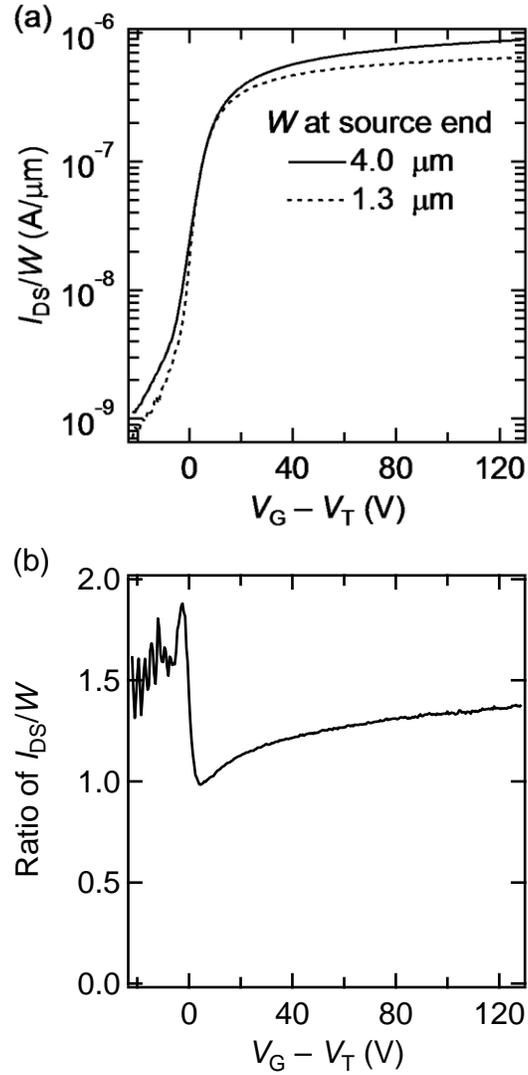

FIG. 5. Transfer characteristics of FETs based on a thin MoS$_2$ flake with a thickness of 23.5 nm (sample #3 in Fig. 4(a)) measured with $V_{DS}$ = 0.1 V and at 333.3 K. (a) $I_{DS}/W$ curves of the devices with the channel widths at the source end of 4.0 and 1.3 μm for the solid and dashed curves, respectively. (b) The ratio of the $I_{DS}$/W value of the 4.0-μm-wide device to that of the 1.3-μm-wide device.

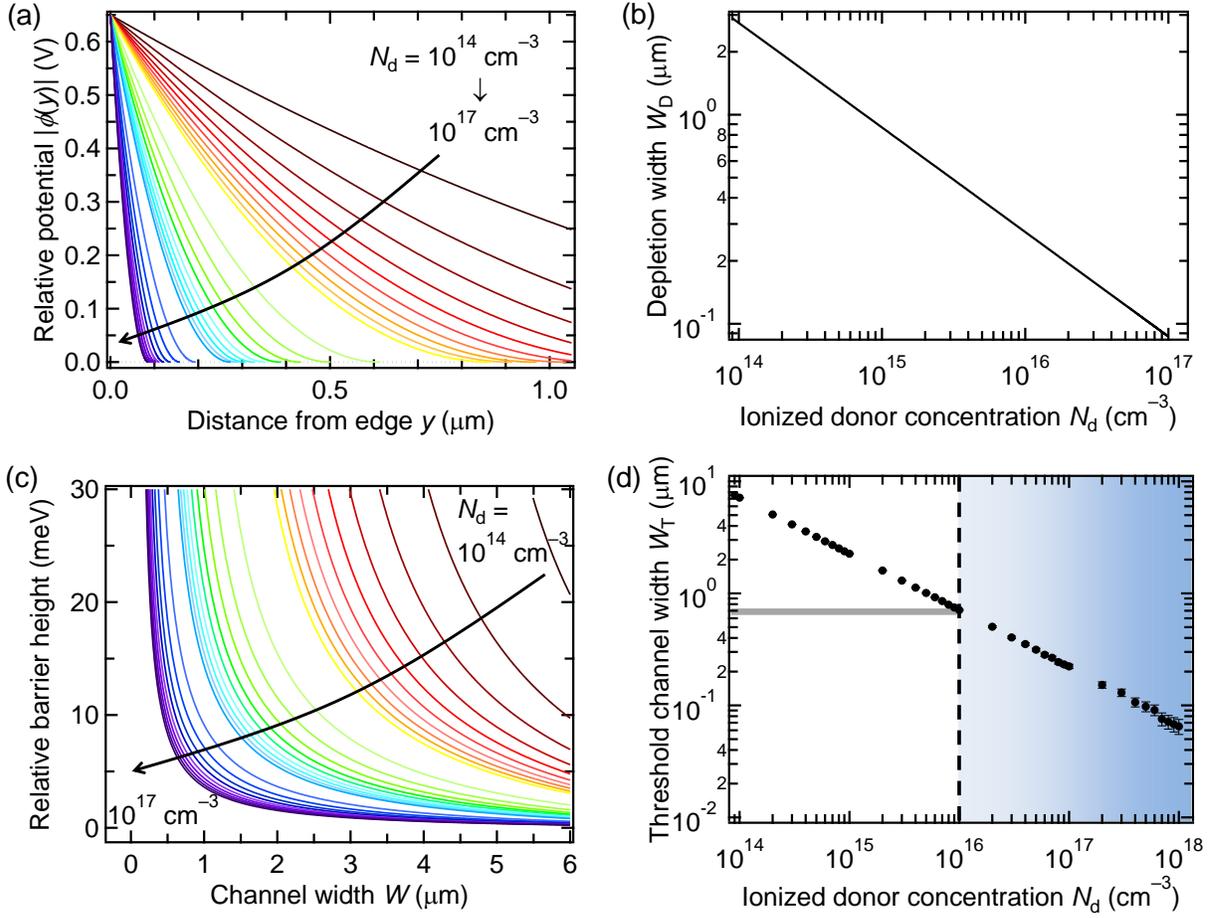

FIG. 6. Theoretical calculation of the effect of the edge-induced band bending with various donor impurity concentrations $N_d$. (a) Variation in the potential $\phi(y)$ in the semiconductor channel. The variable $y$ is the distance from the edge along the channel-width direction. The colors of the plots correspond to $N_d$ = 1 to 10 × $10^{14}$ (black to yellow), 2 to 10 × $10^{15}$ (light green to light blue), and 2 to 10 × $10^{16}$ (blue to purple) $cm^{-3}$. (b) Depletion width $W_D$ of the edge-induced band bending in (a). (c) Schottky barrier heights $\Phi_B^{exp}$. The relative values to the wide-channel limit, $\Delta\Phi_B^{exp} \equiv \Phi_B^{exp} - \Phi_{B\infty}$, are plotted. The colors of the plots are the same as (a). (d) Threshold channel width $W_T$, which was calculated to be equal to the channel width $W$ for $\Delta\Phi_B^{exp} = k_B T$ at 300 K (~26 meV). The gray line indicates the estimated upper limit of $W_T$ without considerable degradation of the charge-carrier mobilities.

25